\documentclass[11pt,twoside]{article}

\usepackage{asp2006}

\usepackage{epsfig}
\usepackage{lscape, amssymb, amsmath}

\newcommand{\Msun}{{M_{\odot}}}

\begin{document}
\title{Promising Observational Methods for Detecting the Epoch of Reionization}   
\author{Matthew McQuinn}   
\affil{Einstein Fellow, UC Berkeley Astronomy Department}    

\begin{abstract} 
It has been several years since the first detection of Gunn-Peterson troughs in the $z > 6$ Ly$\alpha$ forest and since the first measurement of the Thomson scattering optical depth through reionization from the cosmic microwave background (CMB).  Present day CMB measurements provide a significant constraint on the mean redshift of reionization, and the Ly$\alpha$ forest provides a lower bound on the redshift at which reionization ended.  However, no observation has provided definitive information on the duration and morphology of this process.  This article is intended as a short review on the most promising observational methods that aim to detect this cosmic phase transition, focusing on CMB anisotropies, gamma ray burst afterglows, Ly$\alpha$ emitting galaxies, and redshifted 21cm emission. 
\end{abstract}

\section{Introduction}

Reionization is the epoch when essentially all of the intergalactic hydrogen was ionized (which equates to the vast majority of the hydrogen in the Universe at this time; $> 90\%$).  It marks the period when the first galaxies formed, and also when the intergalactic gas was significantly heated to tens of thousands of degrees Kelvin from much smaller temperatures, thereby suppressing the formation of the smallest dwarf galaxies.   At present, we have two constraints on when reionization occurred, which (roughly speaking) show that it happened sometime between the redshifts of $6$ and $15$.  

The first of these constraints comes from the Gunn-Peterson troughs in the Ly$\alpha$ forest (troughs of complete absorption at the redshifted Ly$\alpha$ resonance of intergalactic hydrogen).  These troughs were identified in the spectra of most known $z \gtrsim 6$ quasars (e.g., \citealt{fan06}).  The discovery of these troughs sparked a lot of interest, and many subsequent papers argued that they were evidence that hydrogen reionization ended when the Universe turned a billion years old.  However, the optical depth to redshift through the Ly$\alpha$ resonance is $> 10^5$ in a neutral intergalactic medium (IGM) at $z>6$, making it difficult to establish whether these troughs truly indicated neutral hydrogen fractions near unity.  Very fast temporal evolution in the mean opacity of the Ly$\alpha$ forest is also observed at $z >6$ \citep{fan06}.  This quick evolution is more compelling evidence that the ionization state of the intergalactic hydrogen was changing, but such evolution may still have been possible if reionization were not ending \citep{becker06}.

The second of these constraints comes from observations of the cosmic microwave background (CMB) with the Wilkinson Microwave Anisotropy Probe Satellite (WMAP).  In particular, the large-scale polarization anisotropies were generated when CMB photons scattered off of free electrons generated by reionization.   The WMAP best-fit model to this polarization signal using the fifth year data favors a mean redshift of reionization of $10.4 \pm 1.4$ \citep{komatsu09}.\\

This article is intended as a review of upcoming measurements that have the potential to constrain the reionization era. There are three known techniques that can definitively test whether reionization was occurring by \emph{directly} probing intergalactic neutral fractions of the order of unity:
\begin{enumerate}
\item {\it Thomson scattering off of the free electrons produced by reionization}:  Several percent of the photons originating from before reionization are expected to have been Thomson scattered during this process.  This scattering produced anisotropies in the CMB.
\item {\it Scattering in the wing of the Ly$\alpha$ line of hydrogen}:  Even though the optical depth at line center saturates for neutral fractions of $\sim 10^{-5}$, a photon that started $1000~$km s$^{-1}$ redward of the line center (in the naturally broadened wing of the line) would have experienced optical depths near unity as it traversed a neutral IGM. 
\item {\it Atomic hydrogen's 21cm hyperfine transition}:  Unlike with the Ly$\alpha$ line, only a small fraction of photons that redshifted across this transition would have been absorbed by a neutral IGM.  Thus, this resonance is an unsaturated observable of reionization.  
\end{enumerate}

This review discusses specific applications of the above three techniques and is organized as follows.  Section \ref{sec:model} describes the currently preferred picture (i.e., our best guess) for the reionization process.  Section \ref{sec:CMB}  through Section \ref{sec:21cm} describe respectively how the CMB, gamma ray burst afterglows, Ly$\alpha$ emitter surveys, and redshifted $21$cm radiation can be used as probes of the Epoch of Reionization.


\section{Models for Reionization}
\label{sec:model}

The currently favored model for the reionization history of the intergalactic hydrogen is that stars in the first galaxies produced the ionizing photons that ionized the vast majority of this hydrogen.  It appears that quasars, the most plausible alternative, were not present in sufficient abundance at high redshift to have ionized the IGM (e.g., \citealt{faucher08}).  


The first stars are expected to have formed in $\sim 10^6~M_\odot$ dark matter halos at $z \sim 20-30$.  These halos were just massive enough to excite molecular hydrogen transitions (but not atomic transitions) such that the gas could cool and condense.  However, studies have found that such diminutive halos cannot host sustained star formation: a single supernovae explosion has sufficient energy to unbind all of such a halo's gas (e.g., \citealt{barkana01}), and modest levels of ultraviolet radiation may have been able to dissociate the molecular hydrogen and prevent the halo gas from cooling and forming stars \citep{haiman97}.  Therefore, our best guess is that the Universe had to wait until $\gtrsim10^8~M_\odot$ halos formed for the intergalactic gas to have been reionized.  These were halos massive enough for the gas to cool by transitions from atomic hydrogen and, thereby, host sustained star formation.  In the concordance cosmology, enough of these halos had formed to ionize the intergalactic hydrogen around $z \approx 10$.  Astrophysical uncertainties prevent a precise prediction for when these halos could have reionized the Universe.  However, the exponential growth of collapsed structure in the high-redshift universe makes it difficult for $\gtrsim 10^8~M_\odot$ halos to ionize the Universe significantly earlier than $z=10$ simply because these halos did not exist.  See \citet{barkana01} for a recent review.

The radiation from these early galaxies ionized pockets of the intergalactic hydrogen.  For an ultraviolet spectrum characteristic of any standard stellar initial mass function, the HII-region front width spanned $10$s of comoving kpc at relevant redshifts, which is much smaller than the typical sizes of these regions.  
Sharp ionization fronts resulted in a ``swiss cheese'' topology for the reionization process -- an IGM with fully ionized regions and fully neutral ones, with only a small fraction of the gas at intermediate ionization states.

\begin{figure}
\begin{center}
\includegraphics[height=8.cm]{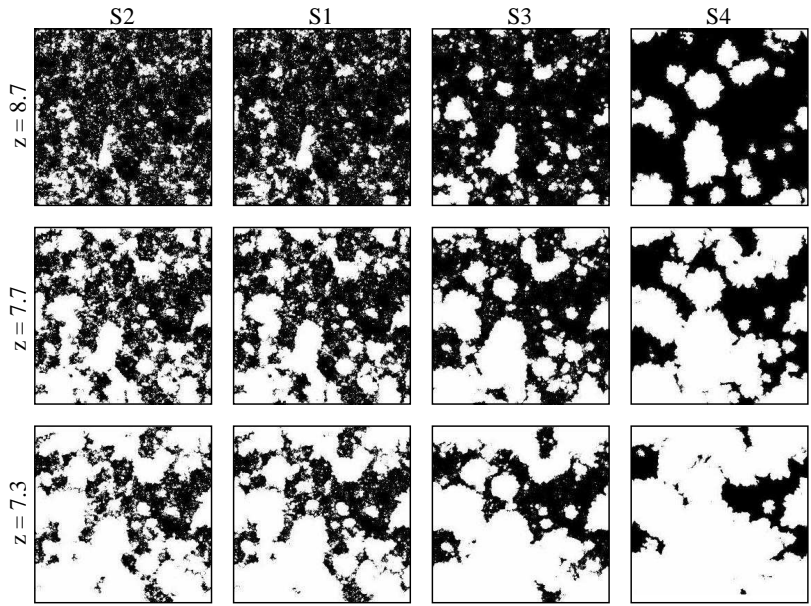}
\end{center}
\caption{Four simulations of reionization from \citet{mcquinn07}.  White regions are ionized and black are neutral.  Each row is compared at the same ionized fraction, and, from left to right, more massive galaxies ionized the simulated intergalactic medium.
\label{fig:morphology}}
\end{figure}

The first models for reionization by stars envisioned that early dwarf galaxies sourced their own small HII regions and that only at the end of reionization did these individual HII regions coalesce.  However, subsequent studies have concluded that, even early on during the reionization process, typical HII regions were sourced by clusters of dwarf galaxies rather than by a single galaxy \citep{furlanetto04a}.  In fact, detailed numerical studies have found that the sizes of the HII regions during reionization would likely have reached tens of comoving Mpc (cMpc) owing to this clustering \citep{iliev05, mcquinn07, Trac07}.  Figure \ref{fig:morphology} shows slices through four simulations of reionization from \citet{mcquinn07} in a $100$ cMpc cube (columns).  The snapshots in each row are compared at the same ionized fraction.  From left to right, less to more massive halos are assumed to have produced the photons that ionized the intergalactic gas.  Because more massive halos are more clustered, the typical HII bubble size also increases from left to right in the Figure.  Note that $\sim 10$~cMpc bubbles are present at the three epochs shown in all four simulations in Figure \ref{fig:morphology}.  

How can we observe reionization and test this simple picture?  The rest of this article focuses on this question.

\section{CMB Anisotropies}
\label{sec:CMB}

The CMB photons decoupled from the primordial plasma $300,000$~yr after the Big Bang and traveled more than ten thousand cMpc in the subsequent $14$~Gyr prior to reaching our telescopes.  Along the way, some of these photons ($\approx 10\%$; \citealt{komatsu09}) scattered off of the free electrons produced during reionization.  Several percent or more of these photons were likely scattered at high redshifts \emph{during} the Reionization Epoch.  There are two important sources of anisotropies that this scattering generated: (1) Polarization anisotropies correlated over $10$s to $100$s of degrees from the scattering of an incident quadrupole temperature anisotropy off of these free electrons \citep{zaldarriaga97}, and (2) kinetic Sunyaev-Zeldovich anisotropies (kSZ), with correlations on $\sim 1-10$~arcminute scales, resulting from the Doppler shifts induced by scattering off of ionized regions with bulk peculiar velocities \citep{sunyaev80}.  

The former source of anisotropy has been used to measure the mean Thomson scattering optical depth through reionization with WMAP, which roughly translates to a measurement of the mean redshift of reionization \citep{komatsu09}.    Observations with the Planck Satellite (which was launched in the summer of 2009) are projected to decrease the uncertainty in the measured mean optical depth value by a factor of $2.5$ \citep{zaldarriaga08}.  Planck may even be able to constrain the duration of reionization in addition to the mean redshift \citep{holder03}.  However, CMB measurements can only constrain the duration at the level of confirming or ruling out relatively long duration reionization episodes ($\Delta z > 5$).  After Planck, observations of the large-scale CMB $E$-mode polarization are projected to be close to their cosmic variance-limit and the constraints on the mean optical depth will be almost saturated (with the error on the optical depth from an ideal experiment improving by no more than another factor of $2$; \citealt{zaldarriaga08}).

The kSZ effect could provide complementary clues into reionization than the large-scale polarization signal.  This anisotropy is second order since it owes to the combination of inhomogeneous free electron density and velocity fields.  The related first order effect due to inhomogeneous velocity and homogeneous electron density is buried under the primary CMB anisotropies.   Once the thermal Sunyaev-Zeldovich anisotropies (owing to scattering in low-redshift hot cluster gas) are removed from a CMB map (which is feasible because of their unique spectral dependence), the kSZ is the dominant source of CMB anisotropies at ${\it l} \gtrsim 4000$.  These ${\it l}$ roughly correspond to several arcminute scales and smaller, just beyond the Silk Damping tail of the primordial anisotropies.  

The kSZ is comprised of two components:  a component due to the density modulation of the free electrons and a component from the patchy distribution of HII regions during reionization (e.g., \citealt{hu00}).  Estimates for the amplitude and shape of the patchy reionization component of the kSZ are model dependent.  Recent numerical simulations of reionization predict that the kSZ anisotropies span a large range of scales with $\delta T \sim 1 ~\mu$K at ${\it l} = 1000-10,000$ \citep{mcquinn05, zahn05a, ilievkSZ}.  The patchy signal will have a different spectral dependence than the kSZ anisotropy owing to density, for which the spectral form is better understood and which is less sensitive to the reionization process.  Furthermore, the amplitude of the patchy kSZ anisotropy is most sensitive to the duration of reionization, and its scale dependence is sensitive to the sizes of the HII regions during reionization \citep{mcquinn05}.   

\begin{figure}
\begin{center}
\includegraphics[height=8.cm, angle=-90]{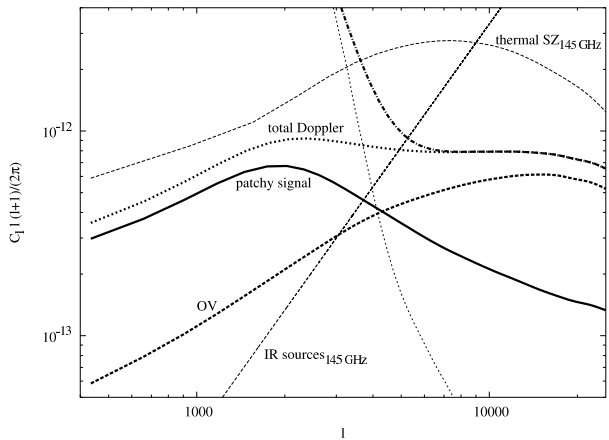}
\end{center}
\caption{Components of the high-{\it l} CMB angular power spectrum from \citet{zahn05a}.  At ${\it l} \lesssim 3000$, the lensed primordial anisotropies dominate the CMB.  At higher {\it l}, the tSZ is most important, but its unique frequency dependence should allow it to be separated from the other anisotropies.  The next most important anisotropy is the kSZ, which has a contribution from scattering off of density fluctuations (the curve labeled ``OV'') and off of large-scale ionization structure from patchy reionization (labeled ``patchy signal'').  The total kSZ signal is labeled ``total Doppler'' and amounts to a $\sim 1 \mu$K signal on several arcminute scales.
\label{fig:patchy}}
\end{figure}

See Figure \ref{fig:patchy} for a breakdown from \citet{zahn05a} of the different anisotropies that contribute to the CMB angular power spectrum at high-${\it l}$, calculated for a specific reionization model.  Two telescopes that are presently in operation, the Atacama Cosmology Telescope (ACT) and South Pole Telescope (SPT), may provide (at best low-significance) detections of the kSZ in the near future.  Recently these telescopes reported their first year results.  Interestingly, the tSZ signal appears to be much smaller than initially projected (and significantly smaller than in Figure \ref{fig:patchy}), such that SPT finds that the kSZ may comprise as much as $\sim 50\%$ of the anisotropies at ${\it l} \sim 4000$ \citep{leuker09}.  This suggests that the kSZ will be easier to measure than previously thought.


A drawback with using the anisotropies in the CMB to study reionization is that the CMB provides only a redshift-integrated measure of the reionization process, making it difficult to pull out information about the structure of reionization at different redshifts.  However, a potential advantage is that all other observational probes of reionization become much more challenging with increasing reionization redshift.

\section{Gamma Ray Bursts}
\label{sec:GRB}

Gamma ray bursts (GRBs) are the most luminous objects at high redshift.  Because higher redshift GRB afterglows tend to be observed earlier in the frame of the GRB (when they are brighter) owing to cosmological time dialation, this bias results in the afterglow flux of GRBs being largely independent of redshift.  As a result, GRB afterglows are detectable with present day telescopes from essentially all redshifts \citep{lamb01, ciardi99}.  In the last several years, the Swift satellite, along with ground-based follow-up programs, have identified three GRBs that originated from $z > 6$, including the present-day record holder for the highest redshift spectroscopically confirmed object, GRB090423 from $z = 8$ \citep{salvaterra09}.    Two proposed space missions, JANUS and EXIST, aim to discover many tens to hundreds of $z > 6$ bursts.

Intergalactic Ly$\alpha$ absorption in the afterglow spectrum of a $z >6$ GRB can be used to study cosmological reionization.  A neutral IGM attenuates the radiation redward of the rest-frame Ly$\alpha$ line in a GRB afterglow by the optical depth \citep{miralda98, mcquinn08}
\begin{equation}
\tau_{\alpha}(\Delta \nu, R_b, \bar{x}_H)  \approx 900 \;{\rm km \, s^{-1}} \; \bar{x}_H \,
\left(\frac{1+z}{8}\right)^{3/2} \; \left(H(z) \, R_b + c \,\frac{\Delta \nu}{\nu} \right)^{-1},
\label{eqn:tau}
\end{equation}
where $\Delta \nu$ is the frequency offset from the Ly$\alpha$ resonance, $R_b$ is the line-of-sight proper size of the host galaxy's intergalactic HII region, $z$ is the redshift of the GRB, and $\bar{x}_H$ is the neutral fraction outside of the HII region.  This broad absorption feature owes to the naturally broadened wing of the Ly$\alpha$ line.  Note that $\tau_\alpha(0, 1 \;{\rm proper~Mpc}, \bar{x}_H) \approx \bar{x}_H$ and that $\tau_\alpha(10,000~{\rm km~s}^{-1}, 1 \;{\rm proper~Mpc}, \bar{x}_H)  \approx 0.1\, \bar{x}_H $.

\begin{figure}
\begin{center}
\plottwo{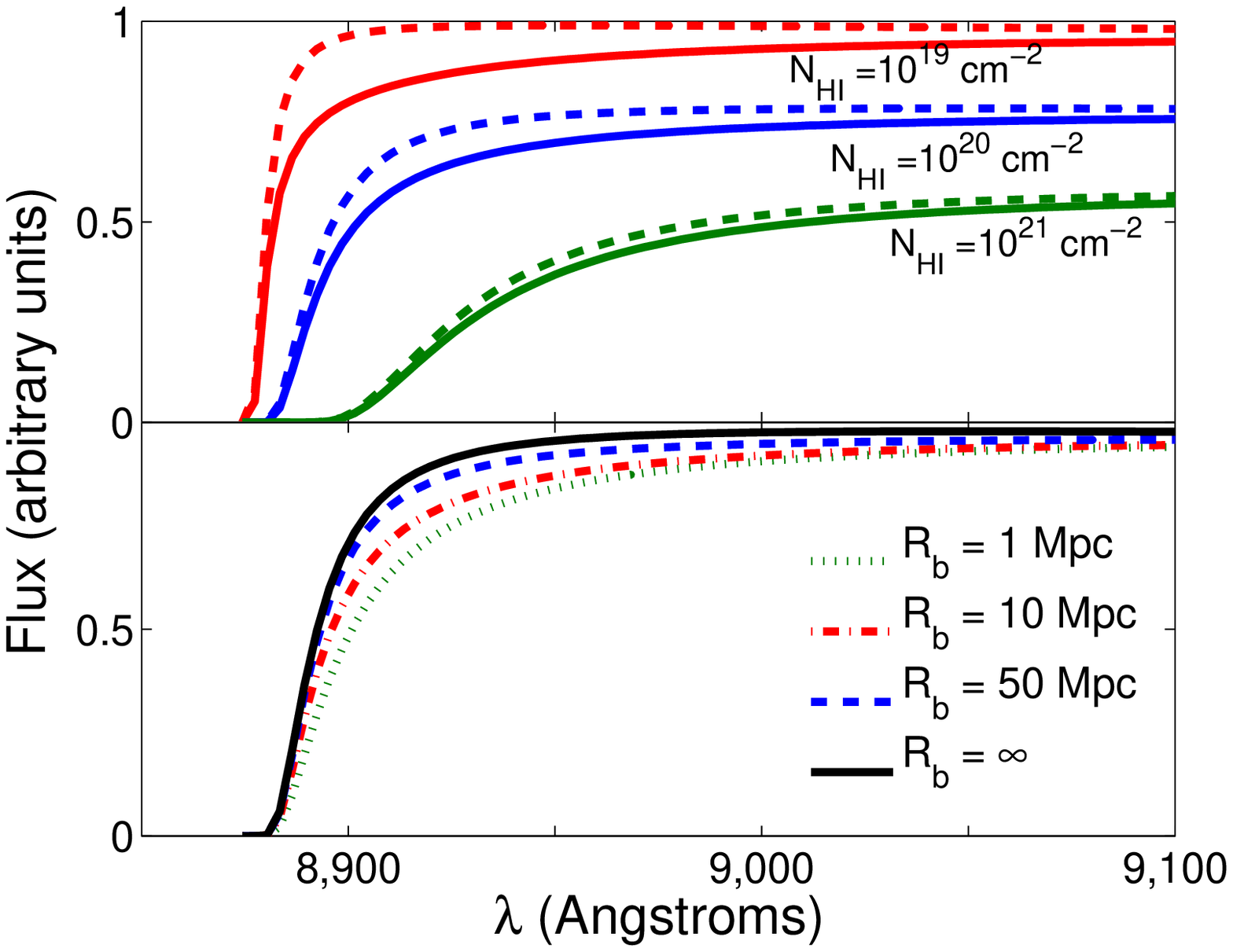}{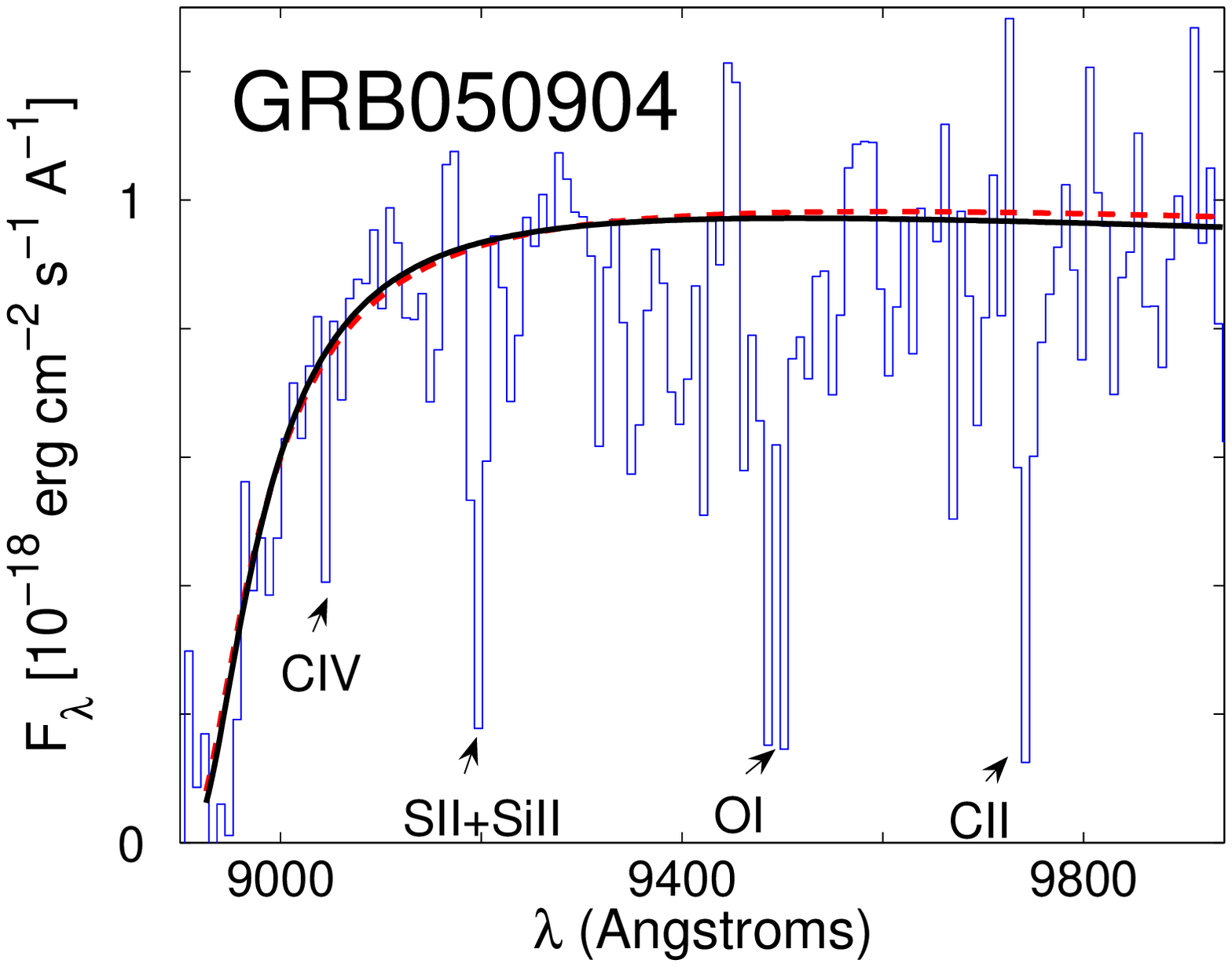}
\end{center}
\caption{{\bf Left}: Example of the GRB afterglow flux near the GRB-frame
Ly${\alpha}$ wavelength.  The top panel depicts the afterglow flux for
both an ionized IGM (dashed curves) and one that is half ionized
where the GRB sits in a $10$ cMpc bubble (solid curves).  Both models include absorption from the host galaxy with column density $N_{\rm HI}$.  The bottom panel illustrates the
effect of the host bubble comoving size on the afterglow flux,
assuming the Universe is half ionized and $N_{\rm HI} = 10^{20} ~ {\rm
  cm}^{-2}$.  {\bf Right}:  The afterglow spectrum of the $z = 6.3$ burst
GRB050904 \citep{kawai06, totani05}, as well as
two best-fit spectral models:  A model that only includes absorption
from the host galaxy (solid curve), and a model that also
includes the absorption from a neutral IGM (dashed curve).  Both models require  $N_{\rm HI}
= 10^{21.6} ~ {\rm cm}^{-2}$, which obscures any IGM contribution. 
\label{fig:dampingwing}
\label{fig:GRB050904}}
\end{figure}

This absorption feature is distinctive from the absorption due to the GRB host galaxy, which all GRB afterglows experience to varying degrees. The attenuation from a neutral IGM scales as $\tau_\alpha \sim \Delta \nu^{-1}$, compared to $\tau_\alpha \sim \Delta \nu^{-2}$ for hydrogen from within the host galaxy.  Therefore, in principle the two can be distinguished \citep{miralda98}.  
The left panels in Figure \ref{fig:dampingwing} compare the shape of the damping wing absorption during reionization for different host-galaxy neutral hydrogen column densities ($N_{\rm HI}$) and ionized bubble sizes within which the GRB host galaxy sits.

In order to detect intergalactic neutral gas with a GRB afterglow, it requires a high signal-to-noise ratio ($S/N$) spectrum to be able to rule out all models where host galaxy absorption was the cause of the absorption feature.  \citet{mcquinn08} estimated for host galaxy neutral columns of $N_{\rm HI} \lesssim 10^{20}\; {\rm cm^{-2}}$, a spectrum with $S/N \gtrsim 10$ per spectral pixel at a resolution of ${\it R} = 3000$ is required.  Such $S/N$ ratios at these moderate resolutions would likely require a ground-based observatory to take a high-redshift GRB's spectrum relatively soon after the prompt $\gamma$-ray emission or that there be followup with a future infrared space telescope such as JWST.   At larger $N_{\rm HI}$, an even higher $S/N$ ratio is needed, and \citet{chen07} find that $\approx 20\%$ of all $z>2$ GRBs have $N_{\rm HI} < 10^{20}\; {\rm cm^{-2}}$.

 The optical afterglow of the $z=6.3$ burst, GRB050904, was so luminous that the minimum S/N ratio and resolution requirements to have a chance at detecting a neutral IGM were almost met even though the afterglow spectrum was taken $3.4$ days after the prompt $\gamma$-ray emission, after the afterglow had dimmed significantly.  Unfortunately for this particular burst, $N_{\rm HI}$ happened to be too large ($\approx 10^{21.6}~$cm$^{-2}$) to constrain the amount of neutral intergalactic gas.  The histogram in the right panel in Figure \ref{fig:dampingwing} is the Subaru spectrum for GRB050904 \citep{kawai06, totani05}.  The solid curve is the best-fit model assuming all of the absorption is from within the host galaxy, which provides an excellent fit to the data.  The afterglows of the two other $z>6$ bursts that have been identified were too faint by the time they were identified as high-$z$ objects to achieve a high enough quality spectrum for this science.

For GRBs occurring at a fixed redshift during reionization, some skewers through the IGM from the GRB will ``see'' much more or much less neutral gas than other skewers because of the inhomogeneous nature of reionization \citep{mesinger08, mcquinn08}.  
\citet{mcquinn08} estimated that the global neutral fraction could be constrained from a single GRB to at best a precision of $\delta x_H \sim 0.2$. 
In addition, the absorption profile can be significantly more complicated than given by equation~(\ref{eqn:tau}) because of the patchy structure of reionization \citep{mesinger08, mcquinn08}.  However, there is not enough information in the Ly$\alpha$ feature to fit a complicated model for the ionization state of the IGM.  \citet{mcquinn08} found that it is only really feasible (even for an extremely large $S/N$ ratio spectrum) to fit for two principle components of the intergalactic absorption, the GRB host HII bubble size and the mean neutral fraction outside of the bubble.

\section{Ly$\alpha$ Emitters}
\label{sec:LAE}

\begin{figure}
\begin{center}
\includegraphics[height=9.cm]{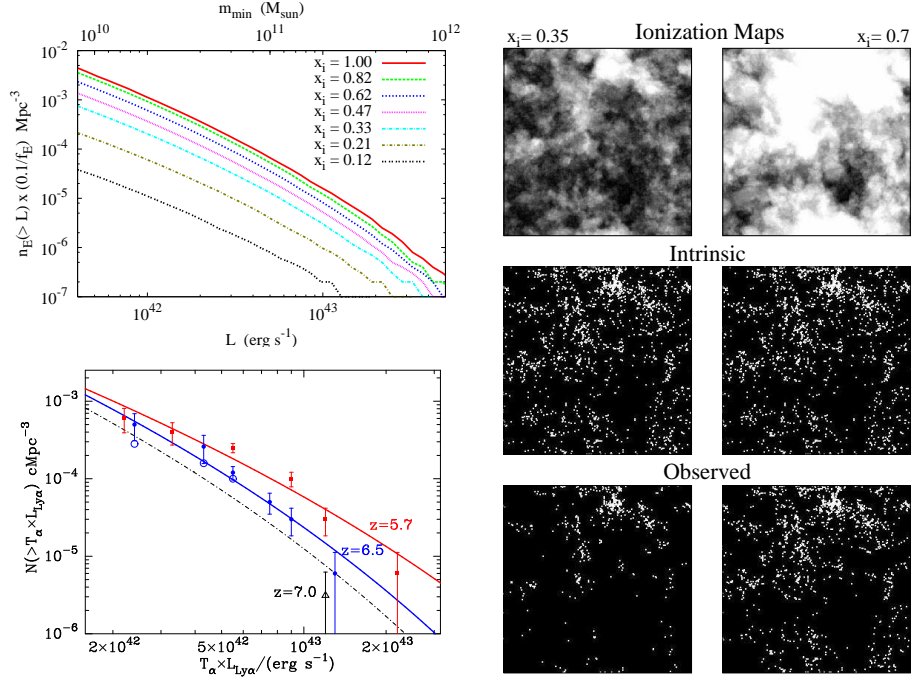}
\end{center}
\caption{Effect of reionization on the statistics of Ly$\alpha$ emitters.  The {\bf top-left} panel presents predictions for the effect of reionization on the $z=6.5$ Ly$\alpha$ emitter luminosity function for different ionization fractions in a particular model for reionization \citep{mcquinn06}.  The $2\times 3$ set of panels {\bf on the right} illustrate the effect that reionization may have on the clustering of emitters.  The top panels in this set show the projected ionization state through the mock survey volume, the middle panels show the intrinsic distribution of the emitters, and the bottom panels show what would be observed.  The observed distribution is significantly modulated by the distribution of neutral gas.  The {\bf bottom-left} panel shows constraints from the Subaru Deep Field on the luminosity function of Ly$\alpha$ emitters at three redshifts, $z = 5.7$ (squares), $z=6.5$ (filled circles), and $z=7.0$ (triangles).  The curves are for a simple model that tries to explain the observed evolution with just evolution in the halo mass function (assuming a one-to-one mapping between halo mass and luminosity), from \citet{dijkstra07}.  The fact that this simple model can fit the data suggests that the observed evolution can be explained without invoking neutral intergalactic gas.
\label{fig:LAE}}
\end{figure}

\begin{figure}
\begin{center}
\includegraphics[width=9.cm]{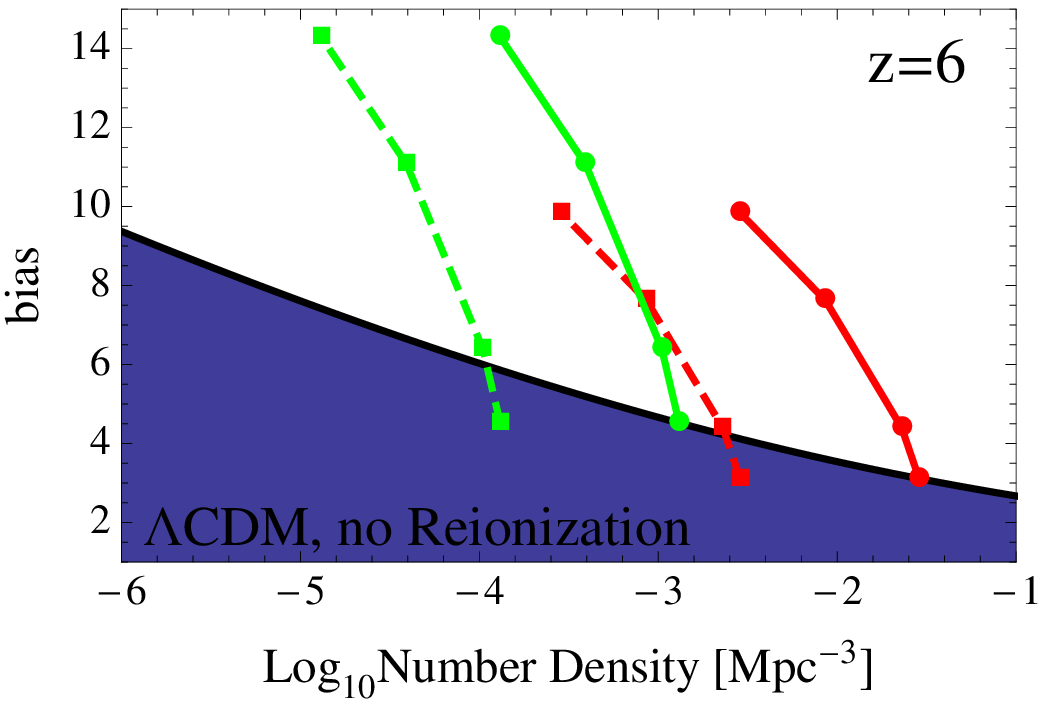}
\end{center}
\caption{Predictions for the large-scale bias of LAEs at $z=6$.  The blue region is the large-scale bias that a population of galaxies at the given number density could achieve if the galaxy clustering were just determined by how dark matter halos cluster.  The points on each curve represent the ``observed'' number density and bias of the LAEs for $x_H = 0, 0.2, 0.5$, and $0.7$, in ascending order.  All of the curves are calculated from the fiducial simulation in \citet{mcquinn07}.  The red curves assume that the galaxies sit in halos with mass $> 10^{10}~\Msun$ and the green curves $> 10^{10.8}~\Msun$.  The solid curves assume that all halos host a LAE, and the dashed assume that only $10\%$ of halos host one.  \emph{During reionization, the clustering of LAEs can be larger than is possible otherwise.}  \label{fig:LAEbias}}
\end{figure}

A reionization-era Ly$\alpha$ emitter that was within a proper Mpc along the line of sight of its HII bubble edge would have had most of its photons scattered in the wing of the Ly$\alpha$ line by the neutral gas at the bubble edge.  However, an emitter located further from the bubble edge than a proper Mpc would have suffered much less attenuation (eqn. \ref{eqn:tau}).  Typical bubble sizes during reionization are conveniently predicted to of the order of a proper Mpc (Fig. 1; \citealt{furlanetto04a, mcquinn06}).  Therefore, Ly$\alpha$ emission from galaxies has the potential to be a useful probe of reionization.

In fact, several narrow-band Ly$\alpha$ emitter surveys at $z > 6$ are underway with the aim of probing this process.  These surveys target gaps in the atmospheric opacity in the infrared to search for strong emission lines, and an emission line in conjunction with no flux blueward of this transition (a Lyman break) is classified as a high-z Ly$\alpha$ emitter candidate.  The brightest of these candidates have been spectroscopically confirmed, which constitutes identifying a broad asymmetric line, characteristic of Ly$\alpha$.  Such follow-up studies have concluded that most of the candidates are real.  The Subaru telescope has been used to identify hundreds of candidates at $z =6.5$ \citep{kashikawa06}, and several at $z = 7.0$ (with $1$ confirmed; \citealt{Iye:2006mb}).  Between $z = 5.7$ and $z=6.5$ (and even between these and $z = 7.0$), significant evolution in the Ly$\alpha$ emitter luminosity function is detected \citep{Iye:2006mb}.  This evolution could be due to intrinsic evolution in the properties of the sources or evolution in the ionization state of the IGM.  It will probably be impossible to distinguish between these two possibilities with the luminosity function alone, but two additional diagnostics may provide insight into the cause of this observed trend.  First, the luminosity function of the continuum redward of Ly$\alpha$ would not have been affected by the ionization state of the IGM, whereas it would have been affected by intrinsic evolution.  Interestingly, in the Subaru data the continuum luminosity function does not decline as significantly between $z=5.7$ and $z=6.6$ as the Ly$\alpha$ luminosity function \citep{kashikawa06}.

The other possibility to distinguish between the two possibilities is through the clustering of Ly$\alpha$ emitters.  Patchy reionization modulates the observed distribution of emitters and can increase the observed clustering significantly.  \citet{mcquinn07} finds in simulations that this can result in the emitters clustering by more than is possible if the emitters were simply biased tracers of the dark matter.  
Predictions for the effect reionization has on the clustering of Ly$\alpha$ emitters are presented in the right panel of Figure \ref{fig:LAE} and in Figure \ref{fig:LAEbias}.  If a sharp increase in clustering is observed, one can in principle perform the following test to convince oneself that it owes to reionization:  Select the high-redshift galaxies with a different technique than by their Ly$\alpha$ emission, such as by the presence of a Lyman break or by H$\alpha$ emission.  Then, measure the flux in Ly$\alpha$ from the selected galaxies.  During reionization, the Ly$\alpha$ emitting subsample could be much more clustered than the full sample of galaxies.  This is in contrast to lower redshifts where Ly$\alpha$ emitters are typically the youngest, least clustered objects.

A common criticism of using the Ly$\alpha$ line to study reionization is that galactic physics, which determines the emergent line shape, is messy.  In particular, the Ly$\alpha$ line is often associated with systematic offsets from the systemic redshift of the galaxy owing to kinematical or radiative transfer effects.   However, the damping wing suppression of the line due to a neutral IGM is fairly robust to these astrophysical effects because this suppression is relatively uniform over $1000~$km~s$^{-1}$ redward of the line center (equation \ref{eqn:tau}).  Astrophysical effects are typically observed to shift the line by several hundred km s$^{-1}$ at $z \sim 3$ \citep{shapley03}, but these shifts could be smaller at high redshift where the emitting halos are substantially less massive.  \citet{mcquinn07} found that shifting the Ly$\alpha$ line of their simulated emitters by $400$~km~s$^{-1}$ to the red did not have a large impact on the properties of the LAEs.

\section{21cm Emission}
\label{sec:21cm}

\begin{figure}
\begin{center}
\includegraphics[width=14.cm]{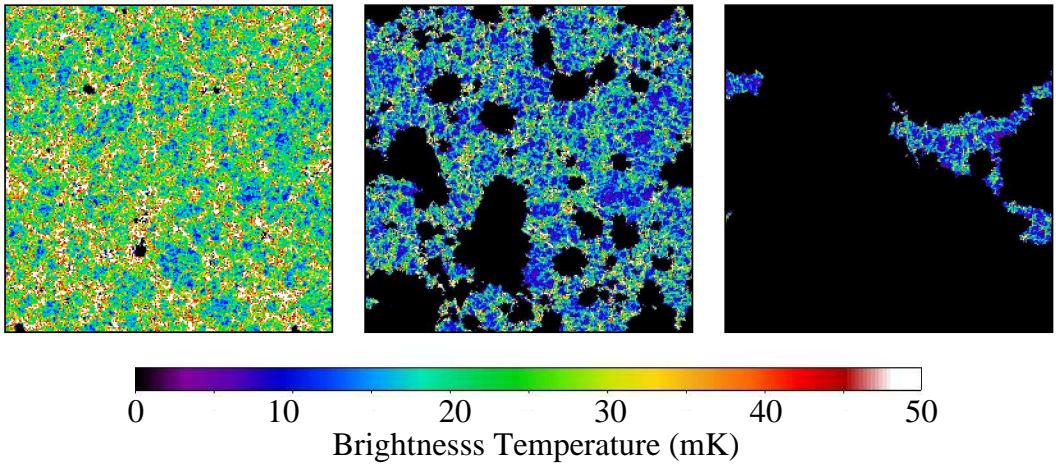}
\end{center}
\caption{Simulated $21$cm emission signal at the beginning, middle, and end of reionization, assuming $T_S \gg T_{\rm CMB}$.  Each panel is $100$~cMpc on a side (and would subtend $\sim 0.6^\circ$ on the sky) and is calculated from a simulation of hydrogen reionization presented in \citet{mcquinn07}.  Interferometric observations in the foreseeable future will produce maps of much lower quality than these simulated maps, but the angular size of the observed maps will typically be much larger.
\label{fig:21cm}}
\end{figure}

Redshifted 21cm emission from the hyperfine line of intergalactic atomic hydrogen has the potential to provide information that is unrivaled by the other probes of reionization. (See \citealt{furlanetto-review} for a recent review.)  Twenty-one~cm absorption studies against high-redshift radio-bright objects are also possible, but likely more challenging (e.g., \citealt{furlanettoforest}).  In principle, 21cm emission can be used to generate a 3-D map of the ionized regions as a function of redshift (frequency).  The excess 21cm brightness temperature over that of the CMB is given by
\begin{equation}
\delta T_{21} = 28 \, x_H  \, \Delta_b \, \frac{T_S - T_{\rm CMB}}{T_S} \left(\frac{1+z}{10}\right)^{1/2}~ {\rm mK},
\end{equation}
where $T_{\rm CMB}$ is the CMB temperature, $x_H$ is the neutral fraction, $\Delta_b$ is the baryon density in units of the cosmic mean, and $T_S$ is the spin temperature (which is a measure of the occupation ratio in the hyperfine excited and ground states).  Theoretical studies find that during the bulk of reionization $T_S$ is likely to be close to the gas temperature and much greater than $T_{\rm CMB}$.  In this limit, $\delta T_{21}$ is independent of $T_S$.   It is expected that spatial fluctuations in $\delta T_{21}$ are dominated by the order unity fluctuations in $x_H$ during most of reionization (at least for spatial scales at which the first generation of instruments are sensitive) and that the volume-averaged signal primarily tracks the ionization history.  Figure \ref{fig:21cm} shows $100~$cMpc slices of the 21cm signal at three times during reionization calculated from a simulation in \citet{mcquinn07} assuming $T_S \gg T_{\rm CMB}$. 

To detect the 21cm signal from $z>6$, this signal needs to be separated from foreground emission that is at least $\sim 10^4$ times larger.  Furthermore, the ratio of the average foreground intensity to that of the 21cm scales as $\sim z^{2.6}$, making 21cm observations increasingly difficult with redshift.  The dominant foregrounds at these frequencies (namely synchrotron and free-free emission) are spectrally smooth, which in principle allows the spatial fluctuations in $\delta T_{21}$ to be separated from these smoother contributions.

\begin{figure}
\begin{center}
\includegraphics[width=8cm]{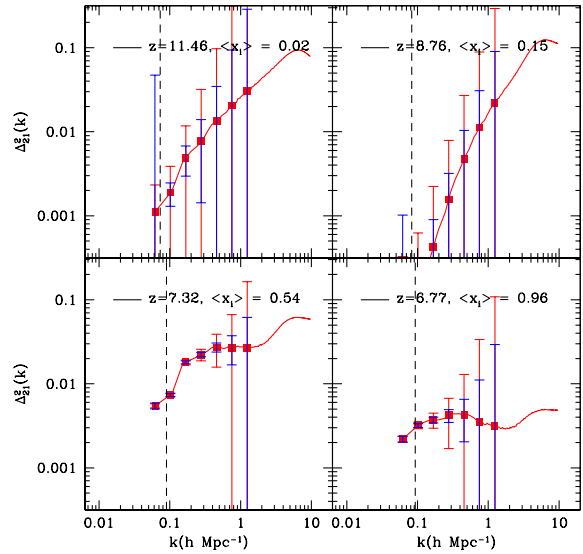}
\end{center}
\caption{Predictions for the 21cm power spectrum at four times during reionization, where $\Delta_{21}^2 \equiv k^3/(2 \pi^2 \, X^2)\, |\delta T_{21}(k)|^2$, $X$ is the global mean of $\delta T_{21}$ if $\bar{x}_H = 1$, and $T_S$ is assumed to be much greater than $T_{\rm CMB}$.  Also included are projections for the constraints that the Murchison Widefield Array can place on this statistic for an integration time of $1000~$hr (outer errorbars), as well as projections for a more centrally concentrated array with the same collecting area and for the same integration time (inner errorbars).  The dashed vertical line is an estimate for the smallest wavevector at which the 21cm power spectrum can be measured.  The signal from smaller wavevectors is likely erased during foreground removal.   From \citet{lidz08}.
\label{fig:21cm_mwa}}
\end{figure}

Several instruments will attempt to integrate down to detect this signal from reionization in the next couple of years.  These efforts can be divided into two categories:  (1) Measurements of the angularly averaged 21cm signal as a function of redshift and (2) interferometric measurements of the spatially fluctuating component of this signal.  The first type -- often referred to as ``Global step'' experiments\footnote{This name originated from the fast step in redshift that the first models of reionization predicted for this global signal.  More recent models predict a more extended step.} -- can be as simple as a well-calibrated single dipole antennae.  Fabulous gain calibration as a function of frequency is essential to detect the global signal, and collecting area is not as pressing of a concern.  Global step experiments are most sensitive to fast reionization scenarios, and they rely on the presumption that the global feature owing to reionization (a $\sim 25~$mK change spanning the duration of this process) has more structure in frequency than the foregrounds.  Of note, the experiment Edges is closest to being sensitive to the global signal, with all systematic trends in their measurement reduced to below $75~$mK once a $\sim 10$th order polynomial is fit and subtracted to remove trends in the instrumental response \citep{bowman08}.  [There are rumors that this $75~$mK number will soon be a factor of $10$ smaller.]   To be sensitive to realistic models of the cosmological signal, \citet{bowman08} estimates that systematics trends need to be below $1~$mK.  

The second category of redshifted 21cm observation includes efforts using the Murchison Wide Field Array (MWA), the Low Frequency Array (Lofar), the 21cm Array, the Precision Array to Map the Epoch of Reionization (PAPER), and the Giant Metrewave Radio Telescope (GMRT).  Each of these interferometers have of order $10^4~$m$^2$ collecting area, and most are built using thousands of dipole antennae.  These searches for the spatially fluctuating signal have the potential to provide much more detailed information about reionization than the Global Step measurements.  Calibration in frequency is less important for these efforts compared to the global step experiments, but the first generation of interferometers (which have typical collecting areas of $\sim 10^4~$m$^2$) require $> 100~$hr integration times to be sensitive to the signal.  In addition, the first generation of these efforts will not be sensitive enough to produce an image of the HII bubbles during reionization and instead must aim for a statistical detection.  

The outer set of errorbars in each panel in Figure \ref{fig:21cm_mwa} represent optimistic estimates for how well a $1000~$hr integration with the MWA can constrain the 21cm power spectrum at different times during reionization, assuming a particular model for this process (from \citealt{lidz08}).  The different panels in this figure correspond to the predicted signal at different redshifts.  The first-generation $21$cm arrays will in general be less sensitive with increasing redshift above $z\gtrsim 6$ because of the increasing brightness temperature of the sky (which sets the instrumental noise floor).  These instruments aim to detect the distinctive redshift evolution in the power spectrum of this signal and to look for the flat power spectrum that is predicted by current models to result at the end of reionization.  These trends can be noted in Figure \ref{fig:21cm_mwa}.

\section{Indirect Methods}
\label{sec:other} 
We have so far concentrated on methods that can directly probe the ionization state of the intergalactic hydrogen.  For completeness, we note two intriguing but more indirect methods.  

First, the ionization state of the hydrogen can be inferred from the absorption of metal ions with similar ionization potentials to hydrogen.  In particular, OI would have been in charge-exchange equilibrium with the hydrogen such that the OI fraction equals the HI fraction \citep{oh02}.  Metal line transitions do not saturate as easily as the HI Ly$\alpha$ line because of their much lower intergalactic abundance.  Observations of several quasar spectra have found a fair sample of OI lines at $z> 6$ (indicating neutral hydrogen) with a  higher abundance than is found at lower redshift \citep{becker06b}.  However, these lines could originate from self-shielding regions and are not necessarily regions characteristic of the IGM.  (It would also be surprising if regions in the IGM that had been enriched with metals were neutral.  It is more likely that the ionization fronts would proceed the enrichment.)

Second, reionization processes are the primary source of heating of the IGM, and they heat it an inhomogeneous fashion (e.g., \citealt{hui03, trac08, furlanetto09}).  It is possible that the signatures of reionization can be detected in measurements of the temperature of the IGM at $z\sim 3-4$ \citep{schaye00, lidz09}.  However, such measurements are difficult, and they require distinguishing the heating due to the ionization of the second electron of helium (which is thought to occur at lower redshift) from that due to the ionization of hydrogen.  If one were to assume that the helium was doubly ionized at the same time that the hydrogen was reionized, the average reionization redshift is constrained to be less than $10$ from the temperature of the $z\sim 3$ IGM \citep{hui03}.  In addition, from comparing the mean transmission in Ly$\alpha$ and Ly$\beta$ forests, \citet{furlanetto09} found evidence for an inverted relationship between the temperature of a gas parcel and its density at $z=6$ (which would indicate that hydrogen reionization ended within $\Delta z \sim 2$ of this redshift).

\section{Conclusions}
We have reviewed the most promising methods to detect reionization.  It is unclear which method will ultimately be the first to detect and characterize this process.  We conclude by summarizing the pros and cons of each method:
\begin{itemize}
\item  Upcoming CMB large-scale polarization measurements with Planck and ground-based experiments should improve the constraint on the mean redshift of reionization by a factor of $2$ to $3$, but can provide only weak constraints on the duration.  In addition, the kSZ anisotropies at ${\it l} \sim 5000$ offer information about the duration and patchiness of reionization, and the South Pole Telescope may offer a tentative detection of the kSZ in the coming years.  Even with a high precision measurement, it will be challenging to interpret what the kSZ signal indicates about reionization.
\item If we are fortunate, in the near future a reionization-era GRB with a small host-galaxy neutral column will be detected.  However, a high signal-to-noise ratio afterglow spectrum at moderate resolution is required to be convinced that the GRB's Ly$\alpha$ damping wing absorption owes to a neutral IGM.   Near infrared spectra of such quality may be possible with fast followup on a bright afterglow or with the next generation of space and ground-based infrared instrumentation.  In addition, a dedicated space telescope could increase the detection rate of $z>6$ bursts by more than an order of magnitude.
\item Narrow band surveys in the near infrared are beginning to obtain large samples of Ly$\alpha$ emitters at $z \gtrsim 6$, and there are some interesting trends in this data.  Evolution in the number counts of these objects alone cannot provide a definitive identification of reionization, but such an identification could be realized in combination with measurements of the Ly$\alpha$ emitter continuum luminosity function or with clustering measurements.  
\item Redshifted 21cm radiation has the potential to provide the most detailed picture of this process, but a measurement of this signal is challenging.  Global step experiments are close to being able to detect a step if reionization were nearly instantaneous, but it is unclear whether this approach can become sensitive to reionization durations that are more theoretically motivated.  Soon several large interferometers will begin searching (and two have already started) for the fluctuating part of the $21$cm signal.  The first generation of interferometers will at best offer a statistically detection of these fluctuations.  However, the second generation of these arrays has the potential to provide detailed maps of the Universe as it became reionized.   
\end{itemize}


\acknowledgements 

I would like to thank the organizers of the Frank N. Bash Symposium for inviting me to give the talk on which this article is based and for organizing this unique symposium.  I would also like to thank Mark Dijkstra, Adam Lidz, and Oliver Zahn for providing figures used in this review.

\bibliographystyle{apj}



\end{document}